\documentclass[article,amsmath,amssymb]{revtex4-2}
\usepackage{graphicx}
\usepackage{dcolumn}
\usepackage[colorlinks]{hyperref}
\usepackage{bm}

\begin{document}

\title{Single Photon Superradiance and Subradiance as Collective Emission From Symmetric and Antisymmetric States}

\author{Nicola Piovella}
\affiliation{Dipartimento di Fisica "Aldo Pontremoli", Universit\`{a} degli Studi di Milano, Via Celoria 16, I-20133 Milano, Italy \&
INFN Sezione di Milano, Via Celoria 16, I-20133 Milano, Italy}
\author{Stefano Olivares}
\affiliation{Dipartimento di Fisica "Aldo Pontremoli", Universit\`{a} degli Studi di Milano, Via Celoria 16, I-20133 Milano, Italy \&
INFN Sezione di Milano, Via Celoria 16, I-20133 Milano, Italy}

\begin{abstract}
Recent works have shown that collective single photon spontaneous emission from an ensemble of $N$ resonant two-level atoms is a rich field of study. Superradiance describes emission from a completely symmetric state of $N$ atoms, with a single excited atom prepared with a given phase, for instance imprinted by an external laser. Instead, subradiance is associated with the emission from the remaining $N-1$ asymmetric states, with a collective decay rate less than the single-atom value. Here, we discuss the properties of the orthonormal basis of symmetric and asymmetric states and the entanglement properties of superradiant and subradiant states. On the one hand, by separating the symmetric superradiant state from the subradiant ones, we are able to determine the subradiant fraction induced in the system by the laser. On the other hand, we show that, as the external laser is switched off and the atomic excitation decays, entanglement in the atomic ensemble appears when the superradiant fraction falls below the threshold $1/N$.
\end{abstract}

\maketitle

\section{Introduction}

Cooperative light scattering by a system of $N$ two-level atoms has been a topic studied since many years~\cite{Lehmberg1970}.  
Many studies in the past have been focused on a diffusive regime dominated by multiple scattering~\cite{Lagendijk1988}, where  light travels over distances much larger than the mean free path. More recently, it has been shown that light scattering in dilute systems  induces a dipole--dipole interaction between atom pairs, leading to a different regime dominated by  single scattering of photons by many atoms. The transition between single and multiple scattering is controlled by the optical thickness parameter {\linebreak} $b(\Delta)=b_0/(1+4\Delta^2/\Gamma^2)$~\cite{labeyrie2003slow,guerin2017light}, where $b_0$ is the resonant optical thickness, $\Delta$ is the detuning of the laser frequency from the atomic resonance frequency and $\Gamma$ is the single atom decay rate. A different cooperative emission is provided by superradiance and subradiance, both originally predicted by Dicke in 1954~\cite{Dicke1954} in fully inverted system. Whereas Dicke superradiance originates from constructive interference between many emitted photons, subradiance is based on destructive interference effect, leading to the partial trapping of light. 
{Subradiant states are important in quantum information to protect entanglement against decoherence, especially in the case of two two-level atoms forming a dark state~\cite{nourmandipour2021entanglement,taghipour2022witnessing}.}

Dicke states have been considered for a collection of $N$ two-level systems~\cite{mandel1995optical}, realized, e.g., by atoms~\cite{gross1982} or quantum dots~\cite{lodahl2004}. In contrast to an initially fully inverted system with $N$ photons stored by $N$ atoms, states with at most one single excitation have attracted considerable attention in the context of quantum information~\cite{brandes2005,karasik2007,pedersen2009}, where the accessible Hilbert space can be restricted to a single excitation by using, e.g., the Rydberg blockade~\cite{tong2004,urban2009,gaetan2009}. A particular kind of single-excitation superradiance has been proposed by Scully and coworkers~\cite{Scully2006,Svidzinsky2008,Svidzinsky2010} in a system of $N$ two-level atoms prepared by the absorption of a single photon (Timed Dicke state).
A link between this single-photon superradiance and the more classical process of cooperative scattering of an incident laser by $N$ atoms  has been proposed by a series of theoretical and experimental papers~\cite{Courteille2010,Bienaime2010,Bienaime2013b}. In such systems of driven cold atoms, subradiance has been also predicted~\cite{Bienaime2012} and then observed~\cite{Guerin2016}; after that, the laser is abruptly switched off and the emitted photons detected in a given direction. Subradiance, by itself, has attracted a large interest for its application in  quantum optics as a possible method to control  spontaneous emission, storing the excitation for a relatively long time.
{For instance, it has been shown~\cite{Scully2015} that it is possible to use subradiance to store a photon in a small volume for many atomic lifetimes and later switch the subradiant state to a superradiant state which emits a photon in a time shorter than the single-atom lifetime. Such a process has potential applications in quantum technology, e.g., quantum memories. Furthermore, it has been proven that the distribution of the atoms in an extended ensemble has a substantial effect on cooperative spontaneous emission~\cite{svidzinsky2015quantum}.  For instance, for atomic distributions with mirror symmetry~\cite{cai2016symmetry}, the anti-symmetric states are subradiant, even when half of the atoms are randomly distributed as long as the mirror symmetry is maintained. Periodic distribution with intrinsic mirror symmetry can be realized in ion traps, and the subradiant states can be prepared by specially tailored antisymmetric optical modes. Also, the presence of an optical cavity may enhance the generation of subradiant states. In ref.~\cite{gegg2018superradiant}, it was shown that by tuning the cavity decay parameter, the nonequilibrium phase transition of cooperative resonance fluorescence changes drastically, amplifying the subradiant Dicke states through cavity-assisted coherences. Letting the system relax into the ground state generates a dark state cascade that can be utilized to store quantum information.}

The aim of this paper is to provide a mathematical description of the single-excitation states in terms of superradiant and subradiant states, i.e., separating the fully symmetric state by the remaining antisymmetric ones. Symmetric and subradiant excited states are distinguished by their decay rates, once populated by a classical external laser and observed; after that, the laser is switched off: the symmetric state has a superradiant decay rate proportional to $N\Gamma$, where $\Gamma$ is the single-atom decay, whereas the antisymmetric states have a decay rate slower than $\Gamma$. 

A crucial point is to determine if such subradiant states are entangled or not, in view of a possible application as quantum memories.
Once  the time evolution of these states is characterized, we will apply the criteria of the spin-squeezing inequalities introduced by T\'{o}th~\cite{toth2007} to detect entanglement in the superradiant and subradiant states. We outline that we limit our study to the linear regime, where the excitation amplitude is proportional to the driving incident electric field. In this linear regime, we must consider
the entanglement criteria which are independent from the value of the driving field, i.e., abandoning these criteria which lead to expressions which depend nonlinearly on the driving field, as  will be discussed in the following. 

The paper is organized as follows. In Section~\ref{s:model}, we present the Hamiltonian describing the dynamics of $N$ two-level atoms interacting with the driving field and write the equation of motion in the linear regime. Then, we calculate the decay rate and the transition rates between different elements of the so-called Timed Dicke basis with its symmetric and antisymmetric states. Section~\ref{s:entanglement} introduces the collective spin operator and the formalism of the spin-squeezing inequalities to assess entanglement. Conclusions are eventually drawn in Section~\ref{s:conclusions}.

\section{The Model}\label{s:model}

\begin{figure}
       {\scalebox{0.6}{\includegraphics{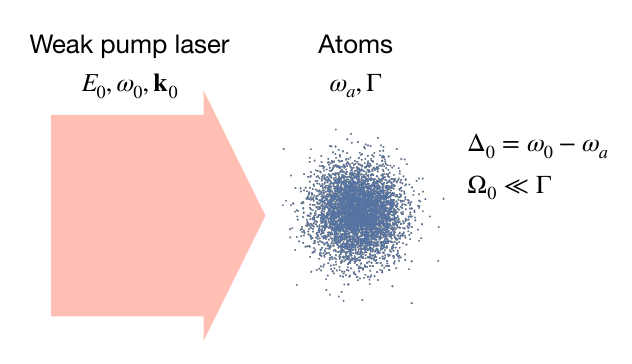}}}
        \caption{Scheme of the system and parameters.}
        \label{fig_setup}
    \end{figure}

We consider $N\gg 1$ two-level atoms with the same atomic transition frequency $\omega_a$, linewidth $\Gamma$ and dipole $d$ (polarization effects are neglected). The atoms are driven by a monochromatic plane wave with electric field $E_0$, frequency $\omega_0$ and  wave vector $\mathbf{k}_0$, detuned from the atomic transition by $\Delta_0=\omega_0-\omega_a$ (see Figure~\ref{fig_setup});  $|g_j\rangle$ and $|e_j\rangle$ are the ground and excited states, respectively, of the $j$-th atom, $j=1,\ldots,N$, which is placed at position $\mathbf{r}_j$.

{We consider the single excitation effective Hamiltonian~\cite{Akkermans2008, Bellando2014} in the scalar approximation. This approximation is appropriate in the dilute limit, where inter-atomic distances are larger than the optical wavelength, making near-field terms decaying as $1/r^3$ negligible. If we assume that only one photon is present, when tracing over the radiation degrees of freedom, the dynamics of the atomic system can be described by the non-Hermitian Hamiltonian $\hat H=\hat H_0-i\hat H_{\rm eff}$, where }\cite{Akkermans2008,Ellinger1994}
\begin{align}
&\hat{H}_0 =-\hbar\Delta_0\sum_{j=1}^N {\hat\sigma_j}^{\dagger}{\hat\sigma_j}+ \frac{\hbar\Omega_0}{2}\sum_{j=1}^N\left(
    \hat\sigma_je^{-i\mathbf{k}_0\cdot\mathbf{r}_j}
    +{\hat\sigma_j}^{\dagger}e^{i\mathbf{k}_0\cdot\mathbf{r}_j}\right) \label{H0}\\
&\hat H_{\rm eff} =\frac{\hbar\Gamma}{2}\sum_{j,m}G_{jm}\,
    \hat{\sigma}_j^\dagger\hat\sigma_m.\label{Heff}
\end{align}

Here,
 $\Omega_0=dE_0/\hbar$ is the Rabi frequency, $\hat\sigma_j=|g_j\rangle\langle e_j|$ and $\hat\sigma_j^\dagger=|e_j\rangle\langle g_j|$ are the lowering and raising operators, {with commutations rules $[\hat\sigma_j,\hat\sigma_m^\dagger]=-\delta_{jm}\hat\sigma_{zj}$, where
{\linebreak} $\hat\sigma_{zj}=|e_j\rangle\langle e_j|-|g_j\rangle\langle g_j|$ are the population difference operators,}  and
\begin{equation}\label{gammajm}
    G_{jm}=
   	\left\{
    \begin{array}{ll}
    \Gamma_{jm}-i\, \Omega_{jm} & \mbox{if}~j\neq m, \\[1ex]
    1 & \mbox{if}~j = m,
\end{array}
	\right.
\end{equation}
where
\begin{equation}\label{gammajm:bis}
    \Gamma_{jm} = \frac{\sin(k_0r_{jm})}{k_0r_{jm}}\quad \mbox{and} \quad
    \Omega_{jm} = \frac{\cos(k_0r_{jm})}{k_0r_{jm}}.
\end{equation}

Note
 that $\hat H_{\rm eff}$ contains both real and imaginary parts, which takes into account that the excitation is not conserved since it can leave the system by emission. By writing the Heisenberg's equations for the operators $\hat\sigma_{j}$ and $\hat\sigma_{zj}$, one can solve them by iteration in powers of $\Omega_0$.  One can see that assuming weak excitation ($\Omega_0\ll\Gamma$), we can approximate (in the Heisenberg's equation for $\hat\sigma_{j}$) $\hat\sigma_{zj}-\approx \hat I_j$,  where $\hat I_j$ is the identity
operator for the $j$th atom~\cite{Bienaime2011}. 
Following the approach reported in refs.~\cite{Akkermans2008,Bellando2014,Ellinger1994}, one can show that this approximation amounts to the linear regime in which all the processes generating more than one photon at the same time are neglected~\cite{Bienaime2011}. Thereafter, only one atom among the $N$ atoms can be found in the excited state, whereas all others are in their ground state. Then, the generic state of the atomic system belongs to an $(N+1)$- dimensional Hilbert space and can be written as~\cite{Scully2006,Scully2015}
\begin{equation}\label{psi}
    |\Psi\rangle=\alpha|g\rangle+\sum_{j=1}^N\beta_j e^{i\mathbf{k}_0\cdot\mathbf{r}_j}|j\rangle
\end{equation}
where $|g\rangle=|g_1,\dots,g_N\rangle$ and $|j\rangle=|g_1,\ldots,e_j,\ldots,g_N\rangle$, $j=1,\ldots,N$. 
{From the Schr\"{o}dinger equation $i\hbar(\partial/\partial t)|\Psi\rangle=\hat H|\Psi\rangle$
and assuming $\alpha\approx 1$ in the  weak-excitation approximation,} we obtain the following equations for the coefficients $\beta_j$ of the state (\ref{psi}):
\begin{align}\label{betaj}
  \dot \beta_j &=
  \left(i\Delta_0-\frac{\Gamma}{2}\right)\beta_j-\frac{i \Omega_0}{2}-\frac{\Gamma}{2}
  \sum_{m\neq j}\widetilde{G}_{jm}\, 
  \beta_m(t), \qquad (j=1,\ldots,N)
\end{align}
where $\dot \beta_j$ refers to the time derivative  of the coefficient $\beta_j$ and $\widetilde{G}_{jm}=G_{jm}e^{-i\mathbf{k}_0\cdot(\mathbf{r}_j-\mathbf{r}_m)}$. {We observe that a more fundamental Master equation approach, describing the atomic system dynamics within the same approximations (scalar approximation and single-excitation approximation), leads to the same Equation (\ref{betaj})~\cite{Ellinger1994,Bienaime2013b}.}

The use of the bare basis $\{|g\rangle,|j\rangle\}$ has the advantage that Equation~(\ref{betaj}) can be easily solved numerically. However, it does not distinguish between symmetric and anti-symmetric states that play a relevant role in our investigation. For this reason we introduce the Timed Dicke (TD) basis with a single excitation~\cite{Svidzinsky2008, Scully2010}:
\begin{equation}
\left\{
 |g\rangle,
 |+\rangle_{\mathbf{k}_0},
 |1\rangle_{\mathbf{k}_0},
 \ldots,
 |N-1\rangle_{\mathbf{k}_0}
\right\},
\end{equation}
where $|g\rangle$ has been introduced above, whereas
\begin{equation}\label{STD}
    |+\rangle_{\mathbf{k}_0}=\frac{1}{\sqrt{N}}\sum_{j=1}^N e^{i\mathbf{k}_0\cdot \mathbf{r}_j}
    |j\rangle
\end{equation}
is the symmetric Timed Dicke (STD) state and
\begin{equation}\label{sk0}
    |s\rangle_{\mathbf{k}_0}=
    \frac{1}{\sqrt{s(s+1)}}
    \left\{
    \sum_{j=1}^{s}
    e^{i\mathbf{k}_0\cdot \mathbf{r}_j}|j\rangle
    - s\, e^{i\mathbf{k}_0\cdot \mathbf{r}_{s+1}}|s+1\rangle
    \right\}, \qquad (s=1,\ldots,N-1)
\end{equation}
are the anti-symmetric ones. The advantage of using the TD basis is that $|s\rangle_{\mathbf{k}_0}$ are collective states involving $s+1$ atoms.  {Considering  (\ref{sk0}) and the definition of $|+\rangle_{\mathbf{k}_0}$, it is straightforward to verify that,} within the considered Hilbert space, the TD basis is complete,~namely:
\begin{equation}\label{unity}
    |g\rangle\langle g|+|+\rangle_{\mathbf{k}_{0}}\langle +|+\sum_{s=1}^{N-1}
    |s\rangle_{\mathbf{k}_{0}}\langle s|=\hat{I}
\end{equation}
and orthonormal, since $\langle g |+\rangle_{\mathbf{k}_0}=\langle g| s\rangle_{\mathbf{k}_0}=
\,_{\mathbf{k}_0}\langle +|s\rangle_{\mathbf{k}_0}=0$, and
$\,_{\mathbf{k}_0}\langle s|s'\rangle_{\mathbf{k}_0}=\delta_{s,s'}$.

To highlight the physical meaning of the TD states, it is useful to evaluate the following transition rates between the basis elements. On the one hand, we find
\begin{align}
\,_{\mathbf{k}_{0}}\langle+|\hat H_0|+\rangle_{\mathbf{k}_0} = -\hbar\Delta_0
\qquad \mbox{and} \qquad
\,_{\mathbf{k}_{0}}\langle +|\hat H_{\rm eff}|+\rangle_{\mathbf{k}_0}   =\frac{\hbar\Gamma_+}{2} - i\hbar\, \Omega_+
\end{align}
with $\hat H_0$ and $\hat H_{\rm eff}$ given in Equations~(\ref{H0}) and (\ref{Heff}), respectively, and
\begin{align}
  \Gamma_+ &=\Gamma\left(1+\frac{1}{N}
  \sum_{j=1}^N\sum_{m=1\atop (m\neq j)}^N\widetilde\Gamma_{jm}\right), \label{G+}\\[1ex]
  \Omega_+ &= \frac{\Gamma}{N}
  \sum_{j=1}^N\sum_{m=1\atop (m\neq j)}^N\widetilde\Omega_{jm} \label{W+}
\end{align}
where
\begin{equation}
\widetilde\Gamma_{jm} = \Gamma_{jm}\cos[\mathbf{k}_0\cdot(\mathbf{r}_j-\mathbf{r}_m)]
\qquad \mbox{and} \qquad
\widetilde\Omega_{jm} = \Omega_{jm}\cos[\mathbf{k}_0\cdot(\mathbf{r}_{j}-\mathbf{r}_{m})]
\end{equation}
and $ \Gamma_{jm}$ and $\Omega_{jm}$ have been introduced in Equation~(\ref{gammajm:bis}). 
For a cloud of cold atoms with a Gaussian distribution with parameter $\sigma_r$, one can prove that $\Gamma_+\approx \Gamma(1+b_0/12)$~\cite{Bienaime2010,Bienaime2011}, where $b_0=3N/\sigma^2$ is the resonant optical thickness, with $\sigma=k_0\sigma_r$ .
Thus, we can conclude that $|+\rangle_{\mathbf{k}_0}$ is a {\it superradiant} state for very large $b_0$~\cite{Svidzinsky2008,Svidzinsky2010}.

On the other hand, the transition rates for the states $|s\rangle_{\mathbf{k}_0}$, $s=1,\dots,N-1$, read
\begin{align}
\,_{\mathbf{k}_{0}}\langle s|\hat H_0|s\rangle_{\mathbf{k}_0} = -\hbar\Delta_0
\qquad \mbox{and} \qquad
\,_{\mathbf{k}_{0}}\langle s|\hat H_{\rm eff}|s\rangle_{\mathbf{k}_0} = \frac{\hbar\Gamma_s}{2} - i\hbar\, \Omega_s 
\end{align}
with
\begin{align}
 &\Gamma_s =\Gamma\left[
  1+\frac{1}{s+1}\left(
  \frac{1}{s}\sum_{j=1}^s\sum_{m=1\atop(m\neq j)}^s\tilde\Gamma_{jm}
  -2\sum_{j=1}^s
  \tilde\Gamma_{j,s+1}\right)
  \right],\label{Gs}\\[1ex]
  &\Omega_s =\frac{\Gamma}{2(s+1)}\left(
  \frac{1}{s}
  \sum_{j\neq m=1}^s\tilde\Omega_{jm}
  -2\sum_{j=1}^s
  \tilde\Omega_{j,s+1}
  \right).
\end{align}

In this case, the decay rates $\Gamma_s$ are less than the single-atom decay rate $\Gamma$, and the states $|s\rangle_{\mathbf{k}_0}$ turn out to be {\it subradiant}~\cite{Scully2015}.

Now we focus on the state (\ref{psi}), which, in the TD basis, reads 
\begin{equation}\label{psi:TD}
    |\Psi\rangle=\alpha|g\rangle+
    \beta_{+}|+\rangle_{\mathbf{k}_0}+\sum_{s=1}^{N-1}\gamma_s|s\rangle_{\mathbf{k}_0},
\end{equation}
with
\begin{align}
  \beta_{+} &=  \,_{\mathbf{k}_0}\langle +|\Psi\rangle=\frac{1}{\sqrt{N}}\sum_{j=1}^N \beta_j,\\
  \gamma_s &= \,_{\mathbf{k}_{0}}\langle s|\Psi\rangle=
  \frac{1}{\sqrt{s(s+1)}}
    \left(
    \sum_{j=1}^{s}
    \beta_j-s\,\beta_{s+1}
    \right).
\end{align}

Thanks to the considerations made above about the properties of the TD states, we can easily find the probability to find our state in a superradiant, i.e.~STD, and subradiant state, that is:
\begin{equation}\label{sup:prob}
P_{+} = |\beta_{+}|^2
\end{equation}
and
\begin{equation}\label{sub:prob}
    P_{\rm sub} = \sum_{s=1}^{N-1}|\gamma_s|^2,
\end{equation}
respectively. Moreover, from the normalization of the state $|\Psi\rangle$, it follows that
\begin{align}
\sum_{j=1}^N|\beta_j|^2 &= |\beta_+|^2+\sum_{s=1}^{N-1}|\gamma_s|^2= P_{+} + P_{\rm sub}.\label{sum_beta2}
\end{align}

Finally,
 the superradiant fraction of excited atoms is given by
\begin{equation}
f_{\rm SR}=\frac{P_{+}}{P_{+} + P_{\rm sub}}
= \frac{\left|\overline{\beta}\right|^2}{~\overline{|\beta|^2}~}\label{fSR}
\end{equation}
where we introduced the mean quantities:
\begin{align}
   &\overline{|\beta|^2} = \frac{1}{N} \sum_{j=1}^N|\beta_j|^2\label{ave1}\\
   &\left|\overline{\beta}\right| = \frac{1}{N}\left|\sum_{j=1}^N\beta_j\right|.\label{ave2}
\end{align}
In turn, the subradiant fraction is $f_{\rm sub}=1-f_{\rm SR}$.

\section{Entanglement Properties of the Superradiant and Subradiant Collective States}\label{s:entanglement}

As the system we are investigating in this paper consists of a large number of atoms, $N\gg 1$, we cannot asses its entanglement properties by individually addressing the single particles. Nevertheless, it is known that spin-squeezing can be used to create large-scale entanglement~\cite{kitagawa1993}. {A two-level atom can be considered a spin-$1/2$ system, described in terms of Pauli's operators of the angular moments}.
Therefore, here we consider suitable generalized spin-squeezing inequalities~\cite{toth2007} based only on collective quantities that are accessible and can be measured experimentally.

Given $N$ two-level atoms, we start defining the following collective angular momentum operators:
\begin{equation}\label{Force-operator}
    \hat J_k=\frac{1}{2}\sum_{j=1}^N \hat\sigma_{j}^{(k)},\qquad\qquad
    (k=x,y,z)
\end{equation}
where $\hat \sigma_j^{(k)}$ are the Pauli matrices associated with the $j$-th atom. If we assume  the state given by Equation~(\ref{psi}), we can calculate the expectation values of the first and second moments of $\hat J_k$ using the result of the previous section (see Appendix~\ref{app:calculations} for  details): 
\begin{eqnarray}
\langle\hat J_x\rangle&=\Re{\rm e}[\alpha^*\sum_{j=1}^N\beta_j],\\[1ex]
\langle\hat J_y\rangle&=-\Im{\rm m}[\alpha^*\sum_{j=1}^N\beta_j],\\[1ex]
 \langle\hat J_z\rangle&= - N \left(\displaystyle
 \frac12 - \overline{|\beta|^2}
 \right),
\end{eqnarray}
and
\begin{align}
    &\langle\hat J_x^2\rangle=\langle\hat J_y^2\rangle=
    \frac{N}{2} \left(
    \frac12 + N \, \left|\overline{\beta}\right|^2 -  \overline{|\beta|^2}
    \right),\\[1ex]
& \langle \hat J_z^2\rangle = N\left[\frac{N}{4}
 - (N-1)\, \overline{|\beta|^2}
\right],
\end{align}
$\beta_j$ being the solutions of Equation~(\ref{betaj}) and $\overline{|\beta|^2}$ and $\left|\overline{\beta}\right|$ are given in Equations~(\ref{ave1}) and (\ref{ave2}), respectively. One can also straightforwardly evaluate the corresponding variances $(\Delta\hat J_k)^2=\langle\hat J_k^2\rangle-\langle\hat J_k\rangle^2$, $k=x,y,z$.

In ref.~\cite{toth2007}, G.~T\'oth and co-workers proved that a sufficient condition to have entanglement is the violation of suitable inequalities involving the first and second moments of the $\hat J_k$ operators evaluated above. Though they proposed four inequalities, in our case, only one of them is relevant to our system, namely:
\begin{equation}
(\Delta\hat J_x)^2+(\Delta\hat J_y)^2+(\Delta\hat J_z)^2 \ge  \frac{N}{2},\label{SS2}
\end{equation}
since the other three T\'oth's inequalities in ref.~\cite{toth2007} are: 
\begin{subequations}
\begin{align}
\langle\hat J^2_x\rangle+\langle\hat J^2_y\rangle+\langle\hat J^2_z\rangle&\le  \frac{N(N+2)}{4},\label{SS1}\\[1ex]
\langle\hat J^2_k\rangle+\langle\hat J^2_l\rangle-\frac{N}{2}&\le  (N-1)(\Delta\hat J_m)^2,\label{SS3}\\[1ex]
(N-1)\left[(\Delta\hat J_k)^2+(\Delta\hat J_l)^2\right]&\ge  \langle\hat J^2_m\rangle+\frac{N(N-2)}{4}\label{SS4}
\end{align}
\end{subequations}
(where $k,l,m$ take all the possible permutations of $x,y,z$) are not useful, as the first and the third are never violated, whereas the second one leads to a condition on the $\beta_j$ going beyond the linear regime assumed to solve Eq.~(\ref{betaj}), 
as shown in Appendix~\ref{app:ineq}.

The inequality (\ref{SS2}) is more interesting. Its l.h.s.~can be rewritten as a function of the superradiant fraction (\ref{fSR}), that is:
\begin{align}
  (\Delta\hat J_x)^2+(\Delta\hat J_y)^2+(\Delta\hat J_z)^2 &=
  \frac{N}{2} + N^2 \overline{|\beta|^2} \left(N\, |\overline{\beta}|^2 - \overline{|\beta|^2}\right)\\[1ex]
  &=\frac{N}{2} + N^2\left(\overline{|\beta|^2} \right)^2 \left(N\,f_{\rm SR} - 1 \right), \label{7b}
\end{align}
which can be solved within the linear regime assumed throughout the paper.
Hence, the inequality 
~(\ref{SS2}) is violated when the superradiant fraction is $f_{SR}<1/N$, thus highlighting the entanglement of the collective atomic state. This will occur when the driving laser is cut and the superradiant component decays faster than the subradiant one, until the subradiant fraction becomes larger than $1-1/N$.

\begin{figure}
       {\scalebox{0.4}{\includegraphics{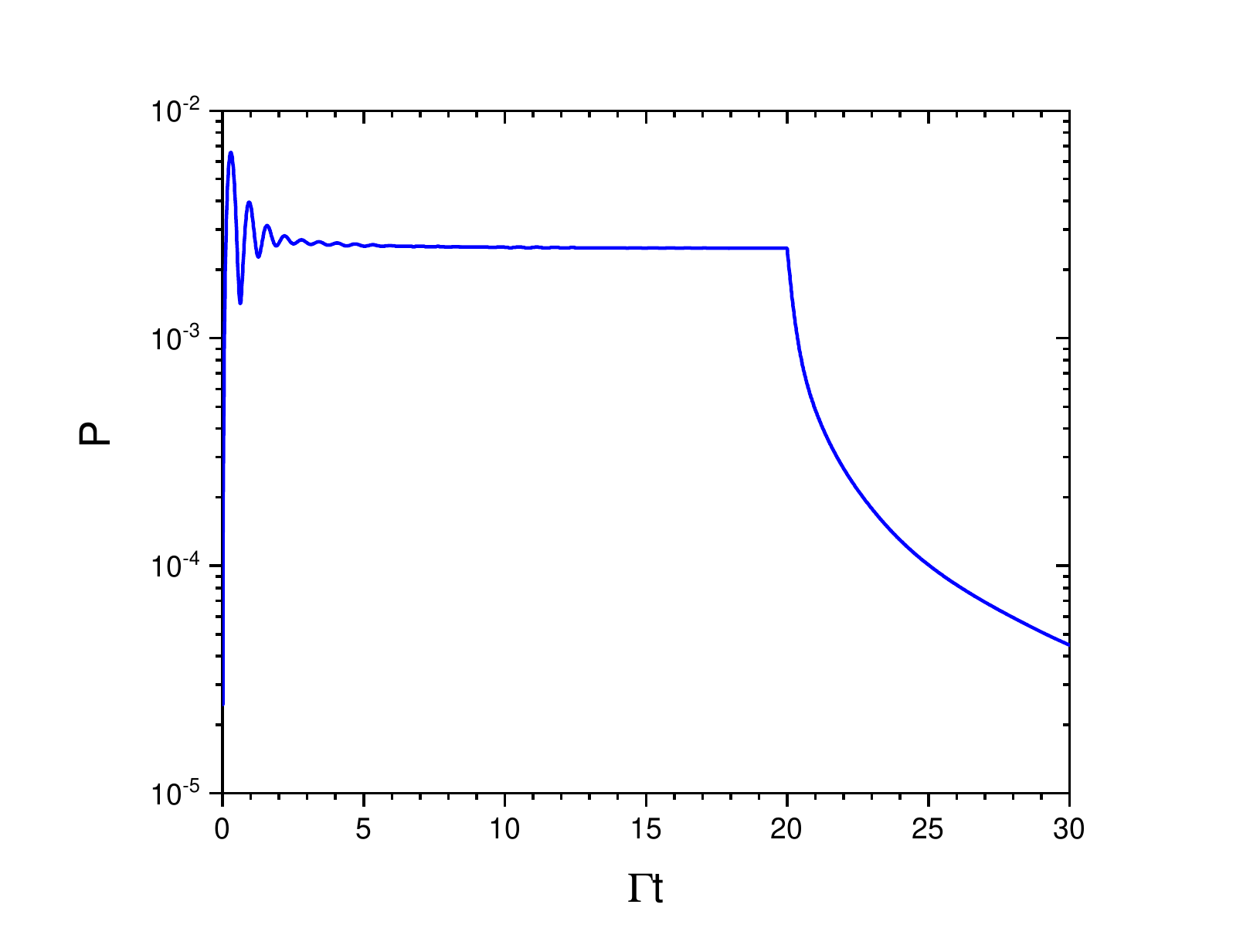}}}
        \caption{Plot of $P=(1/N)\sum_{j}|\beta_j|^2$ vs. time, obtained by solving Equation (\ref{betaj}) for $N=10^3$, $\Delta_0=10\Gamma$ and a Gaussian distribution with $\sigma=k_0\sigma_R=10$. The laser is cut off after $\Gamma t=20$.}
        \label{fig1}
    \end{figure}

\begin{figure}
       {\scalebox{0.4}{\includegraphics{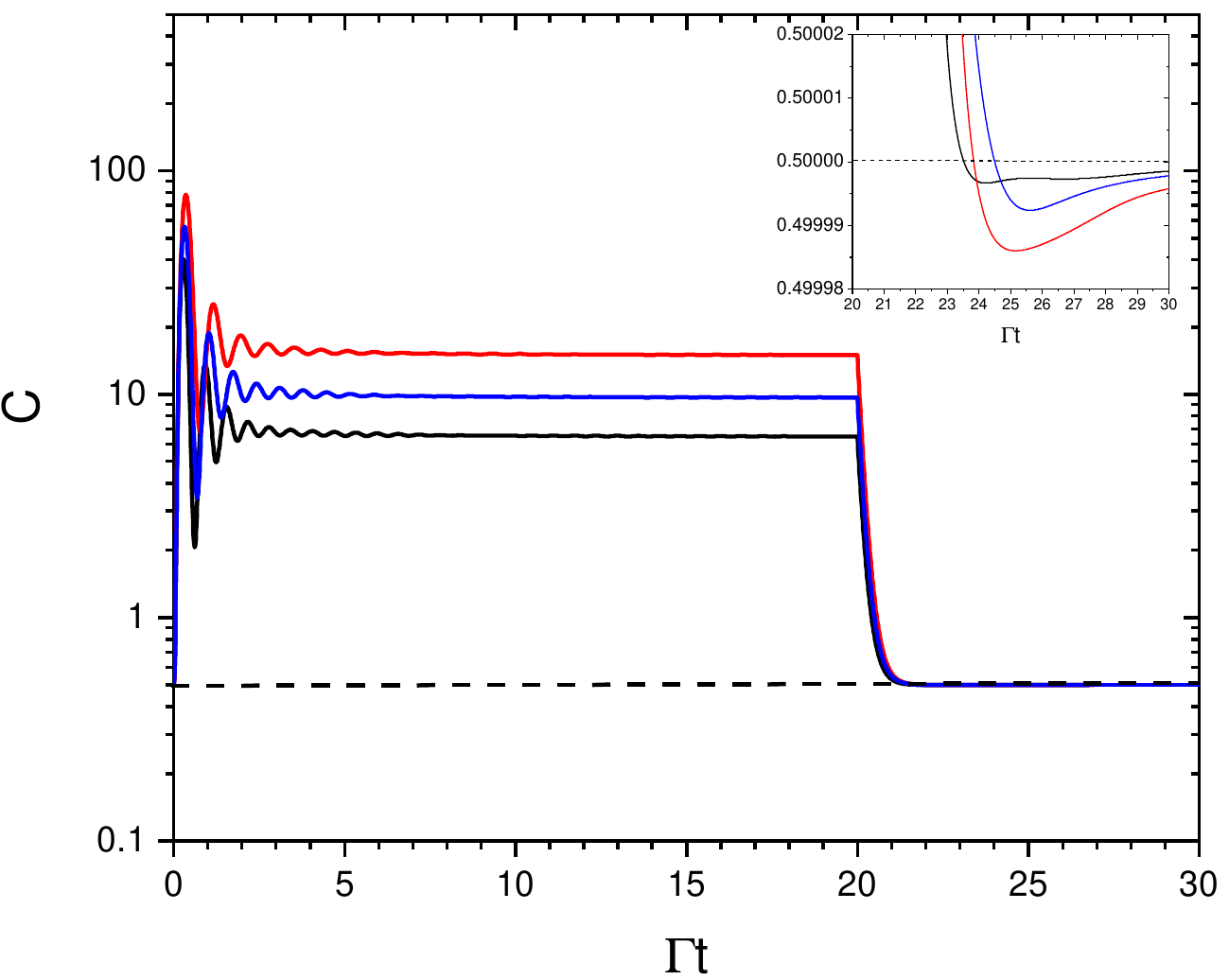}}}
        \caption{$C=[(\Delta\hat J_x)^2+(\Delta\hat J_y)^2+(\Delta\hat J_z)^2]/N$ vs. time for the same parameters of Figure~\ref{fig1} and three different values of detuning, $\Delta_0=8\Gamma$ (red line), $\Delta_0=9\Gamma$ (blue line) and $\Delta_0=10\Gamma$ (black line). The inset magnifies the plot around $C=1/2$, showing the violation of the spin-squeezing inequality (\ref{SS2}).}
        \label{fig2}
    \end{figure}

\begin{figure}
       {\scalebox{0.4}{\includegraphics{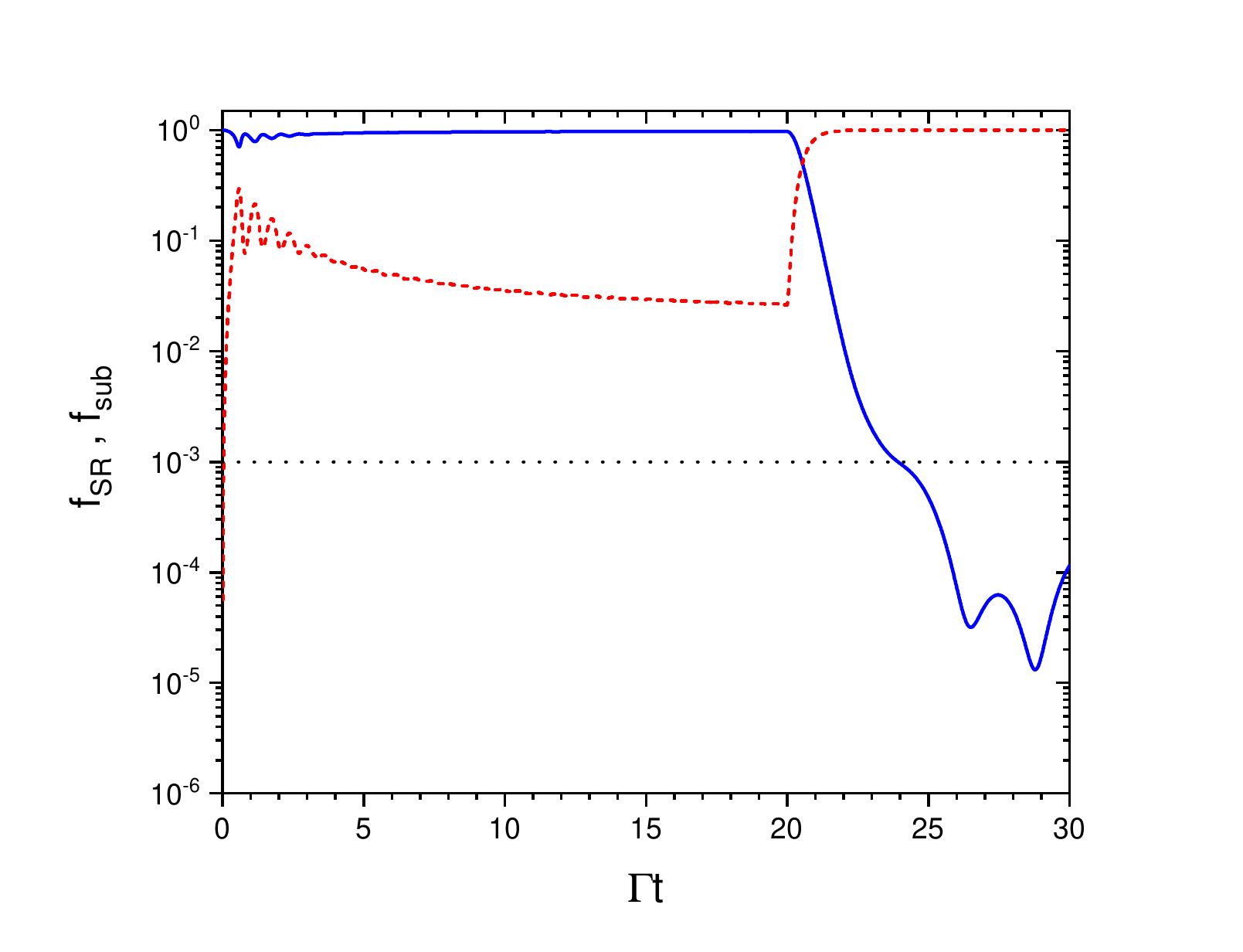}}}
        \caption{Superradiant and subradiant fractions, $f_{SR}$ (continuous blue line) and $f_{sub}$ (dashed red line), for the same parameters of Figure \ref{fig1}. The dotted black line is the value $1/N$.}
        \label{fig3}
    \end{figure}

Figures \ref{fig1}--\ref{fig3} present a typical result. We  numerically evaluate the values of $\beta_j$ solving the linear equations (\ref{betaj}) for $N=10^3$ and a Gaussian distribution with $\sigma=k_0\sigma_r=10$. The laser is cut off after $\Gamma t=20$. Figure~\ref{fig1} shows $P=(1/N)\sum_{j}|\beta_j|^2$ for $\Delta_0=10\Gamma$ vs. time. {Figure~\ref{fig2} shows the left-hand side of Equation (\ref{SS2}) normalized to $N$, i.e., $C=[(\Delta\hat J_x)^2+(\Delta\hat J_y)^2+(\Delta\hat J_z)^2]/N$ vs. time, for the same parameters of Figure~\ref{fig1} and three different values of detuning, $\Delta_0=8\Gamma$ (red line), $\Delta_0=9\Gamma$ (blue line) and $\Delta_0=10\Gamma$ (black line). The inset shows the region where the inequality (\ref{SS2}) is violated, namely when $C<1/2$.} 
Figure~\ref{fig3} shows the superradiant and subradiant fractions, $f_{SR}$ (continuous blue line) and $f_{\mathrm{sub}}$ (dashed red line), as defined by Equation (\ref{fSR}), for the same parameters of Figure~\ref{fig1}. The dotted black line is the value $1/N$. We observe that when the laser is on, the subradiant fraction is only about the $3\%$ of the total excitation. As soon as the laser is cut off, the superradiant fraction decays fast and the subradiant fraction increases, becoming dominating for $\Gamma t> 20.5$. The atoms become entangled when $f_{SR}<1/N$, at time larger than $\Gamma t>24$.

  \unskip
     \unskip 

\section{Conclusions}\label{s:conclusions}

{In this work, we addressed the open question concerning the relation between the entanglement of an atomic ensemble and its superradiant and subradiant fraction. To this aim, we considered the state of $N$ two-level atoms with positions $\mathbf{r}_j$ (with $j=1,\dots,N$), where only one is excited among them. Then, we discussed their symmetry properties: the single-excitation Hilbert space can be spanned by a completely  symmetric state (the ``Symmetric Timed Dicke state'') and $N-1$ asymmetric ones, where the first has a superradiant decay rate proportional to $N$, while the others have subradiant decay rates less than the single-atom value. Remarkably, the projection on the symmetric and asymmetric states allows us to calculate the superradiant and subradiant fractions of the ensemble.}

{To address the problem of the birth of the entanglement,} we studied the relevant case of an ensemble of $N$ atoms driven by an external, uniform laser. In the framework of the linear regime, the probability amplitude of excitation is proportional to the incident electric field, and the superradiant fraction is largely dominant, with only a small fraction of atoms in the subradiant states. However, when the laser is cut off, the superradiant fraction rapidly decays to zero, leaving only the subradiant one, as it has been observed in the experiments of ref.~\cite{Guerin2016}.

{In order to investigate the entanglement properties of the atomic ensemble, we exploited suitable spin squeezing inequalities, based on the first and the second order moments of collective spin operators.} We have found that one of these inequalities is violated when the superradiant fraction decreases below the value $1/N$. Therefore, to have entanglement the probability of finding $N$ atoms in the superradiant state must be less than the average probability per atom to be in the excited state. Conversely, when the superradiant fraction is larger than $1/N$, no entanglement can be detected by measuring the moments of the collective spin operators.

\appendix

\section[\appendixname~\thesection]{Calculation of the First and Second Moments of $\hat J_k$}\label{app:calculations}
In this appendix, we explicitly calculate the first and second moments of the collective angular momentum operators used in the text, namely:
\begin{equation}
    \hat J_k=\frac{1}{2}\sum_{j=1}^N \hat\sigma_{j}^{(k)},\qquad\qquad
    (k=x,y,z)
\end{equation}
where $\hat \sigma_j^{(k)}$ are the Pauli matrices associated with the $j$-th atom. It is useful to introduce the  identities $\hat\sigma_j^\dagger=\left(\hat\sigma_j^{(x)}+
i\hat\sigma_j^{(y)}\right)/2$ and $\hat\sigma_j=\left(\hat\sigma_j^{(x)}-
i\hat\sigma_j^{(y)}\right)/2$, which are the raising and lowering operators appearing in Equations~(\ref{H0}) and (\ref{Heff}), and $\sigma_{j}^{(z)}=|e_j\rangle\langle e_j|-|g_j\rangle\langle g_j|$.
We~define
\begin{eqnarray}
\tilde\sigma_j^\dagger = e^{i\mathbf{k}_0\cdot \mathbf{r}_j}\hat\sigma_j^\dagger\,,\qquad
\tilde\sigma_j = e^{-i\mathbf{k}_0\cdot \mathbf{r}_j}\hat\sigma_j
\end{eqnarray}
and the collective operators:
\begin{eqnarray}
    \hat J_-=\sum_{j=1}^N \tilde\sigma_j\,,\quad
    \hat J_+=\sum_{j=1}^N \tilde\sigma_j^\dagger\quad\mbox{and}\quad
    \hat J_z=\frac{1}{2}\sum_{j=1}^N \hat\sigma_j^{(z)}\label{Jop}
\end{eqnarray}
with commutation relations:
\begin{eqnarray}
    [\hat J_+,\hat J_-]=2\hat J_z,\qquad
    [\hat J_z,\hat J_\pm]=\pm 2\hat J_\pm.
    \label{Jcom}
\end{eqnarray}

We can now apply the formalism of ref.~\cite{toth2007}, calculating the first- and second-order moments
of the collective operators $\hat J_x=(\hat J_+ +\hat J_-)/2$ and $\hat J_y=(\hat J_+ -\hat J_-)/(2i)$.
Explicitly,
\begin{eqnarray}
    \langle\hat J_x\rangle&=&\frac{1}{2}\left\{ \langle\hat J_+\rangle+ \langle\hat J_-\rangle\right\},\\
    \langle\hat J_y\rangle&=&\frac{1}{2i}\left\{ \langle\hat J_+\rangle- \langle\hat J_-\rangle\right\},\\
    \langle\hat J_x^2\rangle&=&\frac{1}{4}\left\{\langle\hat J_+\hat J_-\rangle+ \langle\hat J_-\hat J_+\rangle
    +\langle\hat J_+^2\rangle+\langle\hat J_-^2\rangle\right\},\\
    \langle\hat J_y^2\rangle&=&\frac{1}{4}\left\{\langle\hat J_+\hat J_-\rangle+ \langle\hat J_-\hat J_+\rangle
    -\langle\hat J_+^2\rangle-\langle\hat J_-^2\rangle\right\},
\end{eqnarray}
where the expectations are evaluated using the $\beta_j$ as evaluated from  Equation~(\ref{betaj}) (see Section~\ref{s:model}). Since
\begin{align}
    \tilde\sigma_j|\Psi\rangle &=\beta_j|g\rangle\label{s1},\\[1ex]
    \tilde\sigma_j^\dagger|\Psi\rangle &= \alpha e^{i\mathbf{k}_0\cdot \mathbf{r}_j}|j\rangle\label{s2},\\[1ex]
    \hat\sigma_j^{(z)}|\Psi\rangle &=-\alpha |g\rangle+\beta_j e^{i\mathbf{k}_0\cdot \mathbf{r}_j}|j\rangle
    -\sum_{m=1}^N(1-\delta_{jm})\beta_m e^{i\mathbf{k}_0\cdot \mathbf{r}_m}|m\rangle\nonumber\\
    &=-|\Psi\rangle+2\beta_j
    e^{i\mathbf{k}_0\cdot \mathbf{r}_j}|j\rangle
    \label{s3},
\end{align}
we derive
\begin{align}
 \langle\hat J_-\rangle&=\sum_{j=1}^N\langle\tilde\sigma_j\rangle=\alpha^*\sum_{j=1}^N\beta_j\label{J-},\\[1ex]
 \langle\hat J_+\rangle&=\sum_{j=1}^N\langle\tilde\sigma_j^\dagger\rangle=\alpha\sum_{j=1}^N\beta_j^*\label{J+},\\[1ex] 
 \langle\hat J_z\rangle&=\frac{1}{2}\sum_j\langle\hat\sigma_j^{(z)}\rangle=-\frac{N}{2}+\sum_j|\beta_j|^2\label{Jz},\\[1ex]
    \langle\hat J_x\rangle&=\Re\textrm{e}\left[\alpha^*\sum_{j=1}^N\beta_j\right],\\[1ex]
    \langle\hat J_y\rangle&=-\Im\textrm{m}\left[\alpha^*\sum_{j=1}^N\beta_j\right].
\end{align}

Recalling that the system has a single excitation, we have $\tilde\sigma_j\tilde\sigma_m^\dagger|\Psi\rangle=0$ and $\tilde\sigma_j^\dagger\tilde\sigma_m^\dagger|\Psi\rangle=0$, and we obtain:
\begin{equation}\label{J2}
    \langle \hat J_-^2\rangle = \langle \hat J_+^2\rangle=0.
\end{equation}

Furthermore, 
\begin{align}
    \langle \hat J_+\hat J_-\rangle&=\sum_{j,m}\langle\tilde\sigma_j^\dagger\hat\sigma_j\rangle\nonumber\\
    &=\sum_{j,m}\beta_j^*\beta_m=
    \left|\sum_{j=1}^N\beta_j\right|^2 \label{J+-}
\end{align}
and, using the commutation rule (\ref{Jcom}) and Equations~(\ref{Jz}) and (\ref{J+-}), we find
\begin{align}
    \langle \hat J_-\hat J_+\rangle&=\langle \hat J_+\hat J_-\rangle-2\langle \hat J_z\rangle\nonumber\\[1ex]
    &=N+
    \left|\sum_{j=1}^N\beta_j\right|^2-2\sum_{j=1}^N|\beta_j|^2.\label{J-+}
\end{align}

From these results, we can calculate the expectations of $\langle\hat J_k^2\rangle$, $k=x,y,z$, namely:
\begin{align}
    \langle\hat J_x^2\rangle&=\langle\hat J_y^2\rangle=
    \frac{N}{4}+\frac{1}{2}\left|\sum_{j=1}^N\beta_j\right|^2-\frac{1}{2}\sum_{j=1}^N|\beta_j|^2,\\[1ex]
    \langle \hat J_z^2\rangle&=\frac{N^2}{4}-(N-1)\sum_{j=1}^N|\beta_j|^2.\label{z2}
\end{align}

It is remarkable that the second-order moments do not depend on $\alpha$, which appears only in the expression for $\langle\hat J_{x,y}\rangle$.

We are now ready to also evaluate  the variances in the collective angular momentum operators.
Defining $(\Delta\hat A)^2=\langle\hat A^2\rangle-\langle\hat A\rangle^2$, we have:
\begin{align}
    (\Delta\hat J_x)^2&=\frac{N}{4}+\frac{1}{2}\left|\sum_{j=1}^N\beta_j\right|^2-\frac{1}{2}\sum_{j=1}^N|\beta_j|^2
    -\left\{\Re\textrm{e}\left[\alpha^*\sum_{j=1}^N\beta_j\right]\right\}^2\label{varx}\\[1ex]
    (\Delta\hat J_y)^2&=\frac{N}{4}+\frac{1}{2}\left|\sum_{j=1}^N\beta_j\right|^2-\frac{1}{2}\sum_{j=1}^N|\beta_j|^2
    -\left\{\Im\textrm{m}\left[\alpha^*\sum_{j=1}^N\beta_j\right]\right\}^2\label{vary}\\[1ex]
    (\Delta\hat J_z)^2&=\left(\sum_{j=1}^N|\beta_j|^2\right)\left(
    1-\sum_{j=1}^N|\beta_j|^2\right).
    \label{varz}
\end{align}

Notice that
\begin{align}
    (\Delta\hat J_x)^2+(\Delta\hat J_y)^2 = \frac{N}{2}-\left(\sum_{j=1}^N|\beta_j|^2\right)\left(1-\left|\sum_{j=1}^N\beta_j\right|^2\right)
    \label{varxy}
\end{align}
which is independent of $\alpha$.

All the previous results can be simplified by introducing the mean quantities $\overline{|\beta|^2}$ and  $\left|\overline{\beta}\right|$ given in Equations~(\ref{ave1}) and (\ref{ave2}), respectively. The corresponding expressions are directly reported in Section~\ref{s:entanglement}.

\section[\appendixname~\thesection]{Considerations on {Equations}~(\ref{SS3}) and (\ref{SS4})}\label{app:ineq}

In this appendix, we show that, in our case, Equations~(\ref{SS1}), (\ref{SS3}) and (\ref{SS4}) cannot lead to useful conclusions.

In our case, the l.h.s.~of Equation~(\ref{SS1}) reads:
\begin{equation}\label{7a}
   \langle \hat J_x^2\rangle+\langle \hat J_y^2\rangle+\langle \hat J_z^2\rangle = 
    \frac{N(N+2)}{4} - N^2\sigma_\beta^2
\end{equation}
where we defined the particle variance as
\begin{align}\label{sigma}
    \sigma_\beta^2&=\overline{|\beta|^2}-\left|\overline{\beta}\right|^2,\\[1ex]
     &=\frac{1}{N}\sum_{s=1}^{N-1}|\gamma_s|^2.
\end{align}
As one may expect, Equation~(\ref{SS1}) is true, and the equality sign holds for all the symmetric states, with $\sigma_\beta=0$ or, equivalently, $f_{\rm SR} = 1$.

Concerning Equation~(\ref{SS3}), if we study the following expression (similar results can be obtained for the other combination of the involved expectations):
\begin{equation}\label{7c}
    \langle \hat J_x^2\rangle+\langle \hat J_y^2\rangle-\frac{N}{2}\le
    (N-1)(\Delta\hat J_z)^2,
\end{equation}
we derive:
\begin{align}\label{7c3}
  \left|  \overline{\beta} \right|^2 &\le \overline{|\beta|^2} \left[
    1-(N-1)\overline{|\beta|^2} \right] 
\end{align}
or, equivalently:
\begin{equation}\label{7c4}
\sigma_\beta\ge\sqrt{N-1}\left(|\overline{\beta}|^2+\sigma_\beta^2\right).
\end{equation}
We can clearly see that these inequalities are nonlinear in $\overline{|\beta|^2}$. Recalling Equations~(\ref{ave1}) and~(\ref{ave2}), we conclude that they cannot lead to useful conditions, since  the solution of (\ref{7c3}) or (\ref{7c4}) depends on the $\beta_j^2$ and thus on $\Omega_0^2$, whereas the $\beta_j$ have been obtained in the framework of the linear regime with respect $\Omega_0$, see Equation~(\ref{betaj}).

Equation~(\ref{SS4}) is never violated for our system. In fact, for instance, we have:
\begin{eqnarray}
(N-1)[(\Delta\hat J_x)^2+(\Delta\hat J_y)^2] \ge \langle\hat J^2_z\rangle+\frac{N(N-2)}{4}
\end{eqnarray}
which, using Equations (\ref{varxy}) and (\ref{z2}), yields
\begin{align}
N^2
\overline{|\beta|^2}\left|\overline{\beta}\right|^2 > 0.
\end{align}

\bibliography{Bibliography}

\end{document}